# Observation of Intrinsic Half-metallic Behavior of CrO$_2$ (100) Epitaxial Films by Bulk-sensitive Spin-resolved PES

Hirokazu Fujiwara[1], Masanori Sunagawa[1], Kensei Terashima[2], Tomoko Kittaka[1], Takanori Wakita[2], Yuji Muraoka[1,2], and Takayoshi Yokoya[1,2]

[1]*Graduate School of Natural Science and Technology, Okayama University, Okayama 700-8530, Japan*
[2]*Research Institute for Interdisciplinary Science, Okayama University, Okayama 700-8530, Japan*

**Abstract**
We have investigated the electronic states and spin polarization of half-metallic ferromagnet CrO$_2$ (100) epitaxial films by bulk-sensitive spin-resolved photoemission spectroscopy with a focus on non-quasiparticle (NQP) states derived from electron-magnon interactions. We found that the averaged values of the spin polarization are approximately 100% and 40% at 40 K and 300 K, respectively. This is consistent with the previously reported result [H. Fujiwara *et al.*, Appl. Phys. Lett. **106**, 202404 (2015).]. At 100 K, peculiar spin depolarization was observed at the Fermi level ($E_\text{F}$), which is supported by theoretical calculations predicting NQP states. This suggests the possible appearance of NQP states in CrO$_2$. We also compare the temperature dependence of our spin polarizations with that of the magnetization.

**Keywords**
Half-metallic ferromagnet; Bulk-sensitive spin-resolved photoemission spectroscopy; Non-quasiparticle state; Electron-magnon interaction; Spintronics

**Introduction**

Half-metallic ferromagnets exhibit an electronic spin polarization of 100% at $E_\text{F}$ and are expected to be a suitable material for spintronic devices.[1-5] Chromium dioxide, CrO$_2$, was predicted to be a half-metallic ferromagnet theoretically.[6] Subsequently, approximately 100% spin polarization was obtained in point contact Andreev reflection and spin-resolved photoemission spectroscopy (spin-resolved PES) studies.[2,7-10] However spin-resolved PES studies have previously reported insulating-like spectra, which is in contrast to the metallic behavior of its resistivity. It is also known that the surface of CrO$_2$ tends to change to the antiferromagnetic insulator Cr$_2$O$_3$,[11] suggesting that, in the previous spin-resolved PES studies[8,9], the observed dominant electronic states of were actually that of Cr$_2$O$_3$. Recent bulk-sensitive spin-resolved PES using Xe I line ($h\nu$ = 8.44 eV) as an excitation beam observed the predicted half-metallic electronic structure of CrO$_2$: a clear Fermi edge in the majority spin states and no states at the Fermi level ($E_\text{F}$) with an energy gap of 0.5 eV below $E_\text{F}$.[10]

Electron-magnon interactions are expected to cause anomalous many-body effects in half-metallic ferromagnets because of the 100% spin polarization of the itinerant electrons.[4,12] In finite temperatures, interactions between electrons and thermally excited magnons can occur in ferromagnetic metals. In conventional itinerant ferromagnets, the electron-magnon interaction can produce quasiparticles for both spin projections near $E_\text{F}$ because they have a finite density of states (DOS) for both spins at $E_\text{F}$. On the other hand, in half-metallic ferromagnets, it is predicted that *non-quasiparticles* (NQPs) may appear in the minority spin gap although low-energy electron excitations in the minority-spin states are forbidden in the one-particle picture.[4] Unfortunately, the NQP states decrease spin polarization at $E_\text{F}$, which reduces the advantage of half-metals for applications as spintronic devices. Despite the significance of the NQP states in half-metals in regard to their spin-polarized current behavior, only a few experimental investigations on the many-body effects have been reported.[13]

The appearance of NQP states is predicted in $CrO_2$ by calculations based on dynamical mean-field theory (DMFT) [14]. In the calculated DOS, the NQP states exist in the minority spin gap in an energy region of about 100 meV below $E_F$ at 100 K. Although anomalous depolarization indicating the existence of the NQP states has been observed in several magnetoresistance studies of $CrO_2$,[15,16] direct evidence of the NQP states is still lacking.

In this report, we show the temperature dependence of the electronic states and spin polarization of $CrO_2$ (100) films investigated by bulk-sensitive spin-resolved PES measurements. The spin-resolved PES spectrum at 100 K exhibits peculiar spin depolarization near $E_F$. The energy dependence of the spin polarization is quite similar to that of the DMFT calculations which predict the existence of the NQP states.[14] This suggests that the depolarization is attributed to the emergence of the NQP states.

**Experimental**

The $CrO_2$ (100) epitaxial films grown on $TiO_2$ (100) substrates were prepared using a chemical vapor deposition technique.[17] After the synthesis, the $CrO_2$ film was removed from the quartz tube and then immediately placed under ultra-high vacuum for the spin-resolved PES measurements. In this procedure, the $CrO_2$ film was exposed to the atmosphere for about 5 minutes. No cleaning procedures were made in order to prevent the decomposition of the sample surface by annealing or sputtering. The quality of the prepared $CrO_2$ film surface was evaluated by low energy electron diffraction (LEED) and surface-sensitive PES with a synchrotron radiation of 70 eV at HiSOR BL-5.

Our spin-resolved PES measurements were carried out at 40 K, 100 K, 150 K and 300 K in a spin-resolved PES system with the Mott spin polarimeter (Scienta 2D-spin) whose target was a Au polycrystalline film at Okayama University. The effective Sherman function $S_{eff}$ was determined to be 0.1 by comparison of our spin- and angle-resolved PES spectra of the Bi/Si(111) thin film with that of Ref. 18. The monochromated beam of Xe I resonance line ($h\nu = 8.44$ eV; corresponding probing depth ~ 50 Å [19]) was used as the excitation beam to enhance the bulk-sensitivity. The energy resolution was set to 100 meV. The acceptance angle of the analyzer was ±15° along the [001] direction (easy axis) and ±1° along the [010] direction. We magnetized the sample along the magnetic easy axis by bringing a magnet close to the sample and then measured the spin polarization along the magnetized direction. Careful attention has been paid to maintain the direction of magnetization when we remove the magnet away from the sample. The asymmetry including the sensitivity of a pair of the detectors in the Mott spin polarimeter was calibrated by magnetization reversal of the sample. We obtained the spin polarization by following equation: $P=\{\sqrt{(I^+_L I^-_R)} - \sqrt{(I^-_L I^+_R)}\}/\{\sqrt{(I^+_L I^-_R)} + \sqrt{(I^-_L I^+_R)}\}/S_{eff}$, where $I^{+(-)}_L$ and $I^{+(-)}_R$ are intensity observed by left and right channels, respectively, when magnetizing the sample in the "+ (−)" direction along the easy axis.[20] No background subtraction was applied as the data analysis because the background estimated by the intensity above $E_F$ is much smaller than the signal at and below $E_F$.

**Results and Discussion**

To analyze the quality of the $CrO_2$ sample surface we performed surface-sensitive PES and LEED measurements (figure 1). A peak at a binding energy (BE) of 1 eV, marked by the thick black line, and a metallic Fermi edge were observed in the PES measurements, consistent with results from the bulk-sensitive soft X-ray PES studies.[21,22] The LEED pattern shown as an inset exhibits the clear rectangular-like pattern characteristics of the rutile-type (1×1) structure, which confirms the epitaxial growth of the $CrO_2$ film. From these results, we conclude that single-crystalline $CrO_2$ is the dominant component of our film surface, as has been previously reported.[10,17] Nevertheless, contaminants arising from desorption of the oxygen atoms, primarily amorphous $Cr_2O_3$, may partly exist on the surface. This is evidenced by (i) a small structure at 2 eV BE in our PES spectrum, that is a known characteristics feature of the 3*d* band of $Cr_2O_3$,[11] and (ii) the background of the LEED image. However, the contribution of $Cr_2O_3$ to the spin polarization in an energy range between 1 eV and $E_F$ is ignorable because $Cr_2O_3$ has no electronic states in the energy range according to Ref. 11.

Figure 2 shows the temperature dependence of near-$E_F$ spin-resolved PES spectra and the corresponding spin polarizations. At 40 K and 300 K, averaged values of the spin polarization between 600 meV BE and $E_F$, $P$(Ave.), are approximately 100% and 40%, respectively. This is consistent with the results of our previous measurements.[10] The spin polarization at 40 K is in good agreement with that obtained from the LDA calculations, as seen in Fig. 2. It should be noted that the spin polarization at 100 K significantly drops toward $E_F$ from 150 meV BE. This narrow-range depolarization cannot be explained by the LDA calculations, shown by the yellow solid lines in Fig. 2. One of the possible origins of the narrow-range depolarization can be the NQP states in the minority spin gap. The green dotted lines in Fig. 2 show the spin polarization estimated from the DOS obtained within the LSDA+DMFT calculations for 100 K (Ref. 14). The green lines drop toward $E_F$ from 150 meV BE while the yellow lines keep 100% spin polarization. According to Ref. 14, this depolarization seen in the green line is attributed to the NQP states in the minority spin gap. In order to directly compare the spin polarization curve from our experimental data with that from the theoretical calculations, we multiply the DOS obtained within the LSDA+DMFT calculations by the Fermi-Dirac function for 100 K and broaden it by convolving with a 100 meV Gaussian corresponding to the energy resolution of the measurement. The simulated spin polarization of the DMFT calculation is shown by the blue dashed lines in Fig. 2. The spectral shape of our experimental spin polarization at 100 K fits more closely with the blue line than the yellow line. This indirectly suggests the occurrence of the NQP states at 100 K in the $CrO_2$ film.

To validate the appearance of the narrow-range depolarization, we discuss the differences between the temperature dependence of $P$(Ave.) and that of $P(E_F)$. Figure 3 shows a comparison of the temperature dependence of the values of the spin polarization with the normalized magnetization curve from Ref. 17. At 300 K, $P(E_F)$ has the same value as $P$(Ave.). The difference between $P(E_F)$ and $P$(Ave.) is remarkable at lower temperature, showing that the narrow-range depolarization is more considerable at low temperature rather than at room temperature. The reason seems to be related to the suppression of the depolarization within an energy region from $E_F$ to 600 meV BE (wide-range depolarization). Around room temperature, the wide-range depolarization is dominant in the spin-resolved PES measurements so that the contribution of the NQP states is less considerable. In contrast, at low temperature, the wide-range depolarization is strongly suppressed and as a result the narrow-range depolarization, characteristics of the NQP states, is more pronounced.

Lastly, in order to clarify the origins of the wide-range depolarization, we compare $P$(Ave.) with the magnetization curve obtained from the SQUID measurements. $P$(Ave.) shows a temperature dependence similar to that of the magnetization at low temperature, which supports the bulk-sensitivity of our measurements. However, $P$(Ave.) drops more rapidly with increasing temperature than the magnetization around room temperature. Similar behaviors were reported in an X-ray magnetic circular dichroism (XMCD) and spin-resolved PES study of $La_{0.7}Sr_{0.3}MnO_3$ (LSMO).[23] The LSMO study suggests that the more surface-sensitive the measurement, the more rapidly the obtained magnetization curve drops with increasing temperature. This is due to the effects of the surface boundary of the LSMO. The temperature dependence of *P*(Ave.) in Fig. 3 is similar to that of the magnetization curve obtained from the XMCD measurements in the LSMO study, which indicates that our spin-resolved PES measurements are more bulk-sensitive than those in the LSMO study. This is because a Xe I resonance line was used as an excitation beam. Nevertheless, the value of *P*(Ave.) at 300 K is quite different from that of the magnetization, which could be attributed to the surface effect on the $CrO_2$ (100) film based on the LSMO study. Therefore, in order to investigate the behavior of the NQP states, it is sufficient to make *bulk-sensitive* and *low-temperature* (below 150 K) spin-resolved PES measurements on $CrO_2$, because the wide-range depolarization derived from the surface effect is sufficiently suppressed in the low temperature region. The energy dependence of our experimental spin polarization at 40 K is consistent with that of band calculation for $CrO_2$ (100) surface.[25] To understand the temperature dependence of the experimental spin polarization, theoretical calculations for $CrO_2$ (100) surface at finite temperatures is highly desired.

**Conclusion**

We performed low-energy bulk-sensitive spin-resolved PES measurements on half-metallic ferromagnet $CrO_2$ (100) epitaxial films to observe the NQP states. We found that averaged values of the spin polarization were approximately 100% and 40% at 40 K and 300 K, respectively, which is consistent with our previous report. At 100 K, we observed the peculiar spin depolarization in the energy region between 100 meV BE and $E_F$, which can be well explained by a theory predicting NQP states. This suggests the indirect observation of the NQP states. We also found that the depolarization was noticeable below 150 K because the wide-range depolarization derived from the surface effects was suppressed at low temperature. These results show that the bulk-sensitive and low-temperature spin-resolved PES is a promising experimental technique to observe the NQP states.

**Acknowledgments**

We thank Vladimir N. Strocov and Markus Donath for fruitful discussions. We also appreciate S. J. Denholme proofreading the manuscript. We also thank T. Fukura, T. Nagayama, and A. Takeda for their assistance in spin-resolved PES measurements. This work was partially supported by the Program for Promoting the Enhancement of Research Universities and a Grant-in-Aid for Japan Society for the Promotion of Science (JSPS) Fellows (No.16J03208) from the Ministry of Education, Culture, Sports, Science and Technology of Japan (MEXT). LEED and PES measurements at HiSOR were conducted under the project (BL-5-14-A6). H. F. would like to acknowledge the support from the Motizuki Fund of Yukawa Memorial Foundation. H. F. was supported by a Grant-in-Aid for JSPS Fellows.

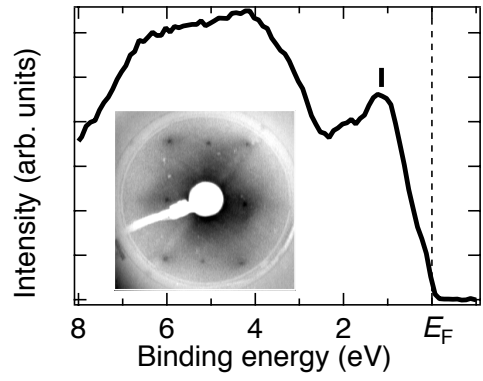

Figure 1  Valence band PES spectrum taken at *hv* = 70 eV at 300 K.  Inset shows the LEED pattern at an incident electron energy of 45 eV.

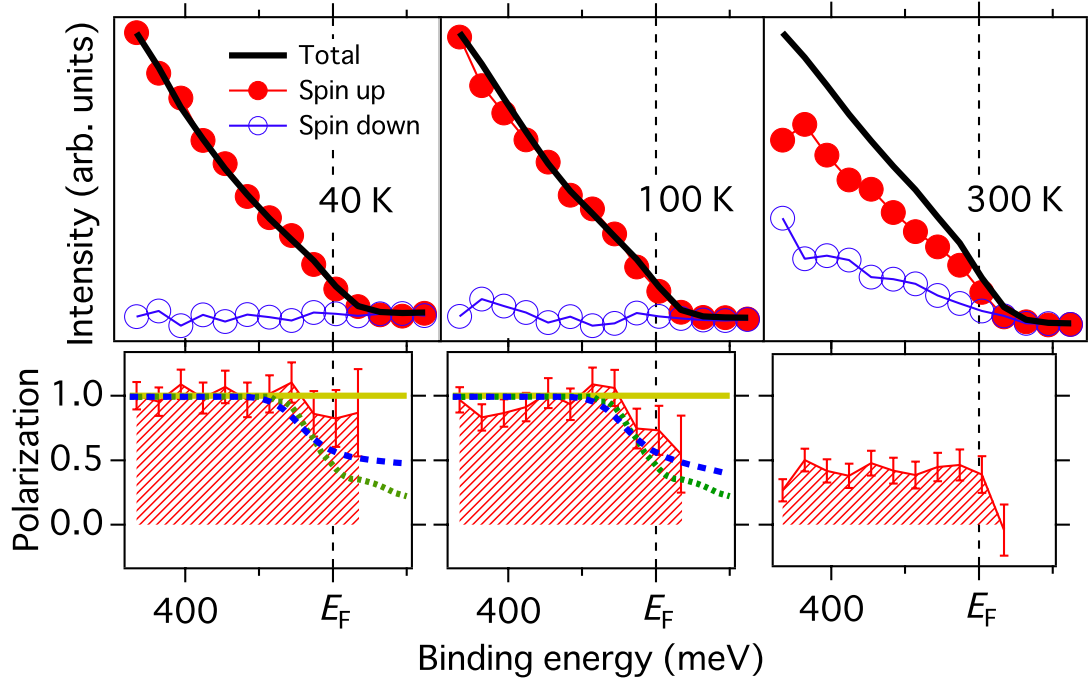

Figure 2 Temperature dependence of near-$E_F$ spin-resolved PES spectra (top) and spin polarization with statistical error bars (bottom) taken at $h\nu = 8.44$ eV. The error bars are estimated from $1/(S_{eff}I^{1/2})$ [24] where $I$ is the total intensity. The yellow solid lines and green dotted lines in spin polarization at 40 K and 100 K show the energy dependence of spin polarization obtained from our LDA calculation and a DMFT calculation at 100 K from Ref. 14. The blue dashed lines show that obtained from the spin dependent DOS multiplied by the Fermi-Dirac function of each temperature and convoluted with a 100 meV Gaussian corresponding to the energy resolution of the measurement.

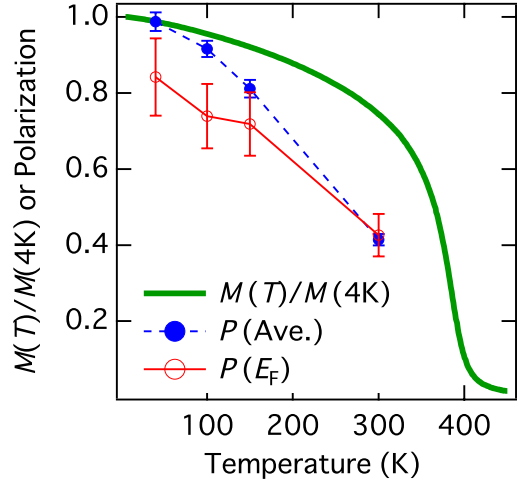

Figure 3 Comparison of the temperature dependence of the spin polarization $P$ obtained by our measurements with the magnetization along $b$-axis direction in an external magnetic field of 0.5 T from Ref. 17. The data points of the spin polarization $P$(Ave.) and $P(E_F)$ represent the average values in the region between 600 meV BE and $E_F$ and between 100 meV BE and $E_F$, respectively. The statistical error bars are estimated from $1/(S_{eff}I_{sum}^{1/2})$ where $I_{sum}$ is the intensity summed in the corresponding averaged energy region.